\documentclass{article}
\usepackage{spconf,amsmath,graphicx}

\usepackage{amsfonts} 
\usepackage{gensymb}  
\newcommand\blfootnote[1]{%
  \begingroup
  \renewcommand\thefootnote{}\footnote{#1}%
  \addtocounter{footnote}{-1}%
  \endgroup
}
\usepackage{xpatch,xcolor}
\usepackage{hyperref}
\hypersetup{
    colorlinks=true,
    linkcolor=purple,
    filecolor=magenta,      
    urlcolor=purple,
}
\usepackage{booktabs}

\title{Global HRTF Interpolation via Learned Affine Transformation of Hyper-conditioned Features}
\name{Jin Woo Lee$^1$, Sungho Lee$^1$, Kyogu Lee$^{1,2,3}$}
\address{
    $^1$Department of Intelligence and Information, Seoul National University \\
    $^2$Interdisciplinary Program in Artificial Intelligence, Seoul National University \\
    $^3$AI Institute, Seoul National University \\
    \texttt{\small\{jinwlee,sh-lee,kglee\}@snu.ac.kr}
}

\begin{document}
\ninept

\maketitle

\begin{abstract}
Estimating Head-Related Transfer Functions (HRTFs) of arbitrary source points is essential in immersive binaural audio rendering.
Computing each individual's HRTFs is challenging, as traditional approaches require expensive time and computational resources, while modern data-driven approaches are data-hungry.
Especially for the data-driven approaches, existing HRTF datasets differ in spatial sampling distributions of source positions, posing a major problem  when generalizing the method across multiple datasets.
To alleviate this, we propose a deep learning method based on a novel conditioning architecture.
The proposed method can predict an HRTF of any position by interpolating the HRTFs of known distributions.
Experimental results show that the proposed architecture improves the model's generalizability across datasets with various coordinate systems.
Additional demonstrations show that the model robustly reconstructs the target HRTFs from the spatially downsampled HRTFs in both quantitative and perceptual measures.
\end{abstract}


\blfootnote{Code and models are available at: \url{https://github.com/jin-woo-lee/hrtf-interpolation}}
\blfootnote{Sound samples are available at: \url{https://bit.ly/3DdmPu9}}

\section{Introduction}
\label{sec:intro}

Humans can localize sound using a variety of perceptual cues, including interaural time difference, interaural level difference, and spectral shifts in the source signal caused by interactions with the pinnae, head, and torso of the listener \cite{wright1974pinna,asano1990role}. 
Head-Related Transfer Function (HRTF) models the spectral changes in acoustic signals as they travel from a sound source to the listener's ears.
Since the anthropometric measurements are different for each listener, measuring HRTFs is essentially an individualized estimation.
Once the HRTFs for a subject are obtained, virtual auditory space can be created by filtering a source sound with the transfer functions that correspond to the subject's spectral cues.
By presenting the filtered sound through headphones, one can give the listener the perception that the sound is localized at the desired virtual space \cite{hendrix1996sense}.
Methods for finding out the HRTFs of each individual have been widely studied \cite{li2020measurement,xu2007individualization}.

\begin{figure}
    \begin{minipage}[b]{1.0\linewidth}
        \centering
        \centerline{\includegraphics[width=0.75\linewidth]{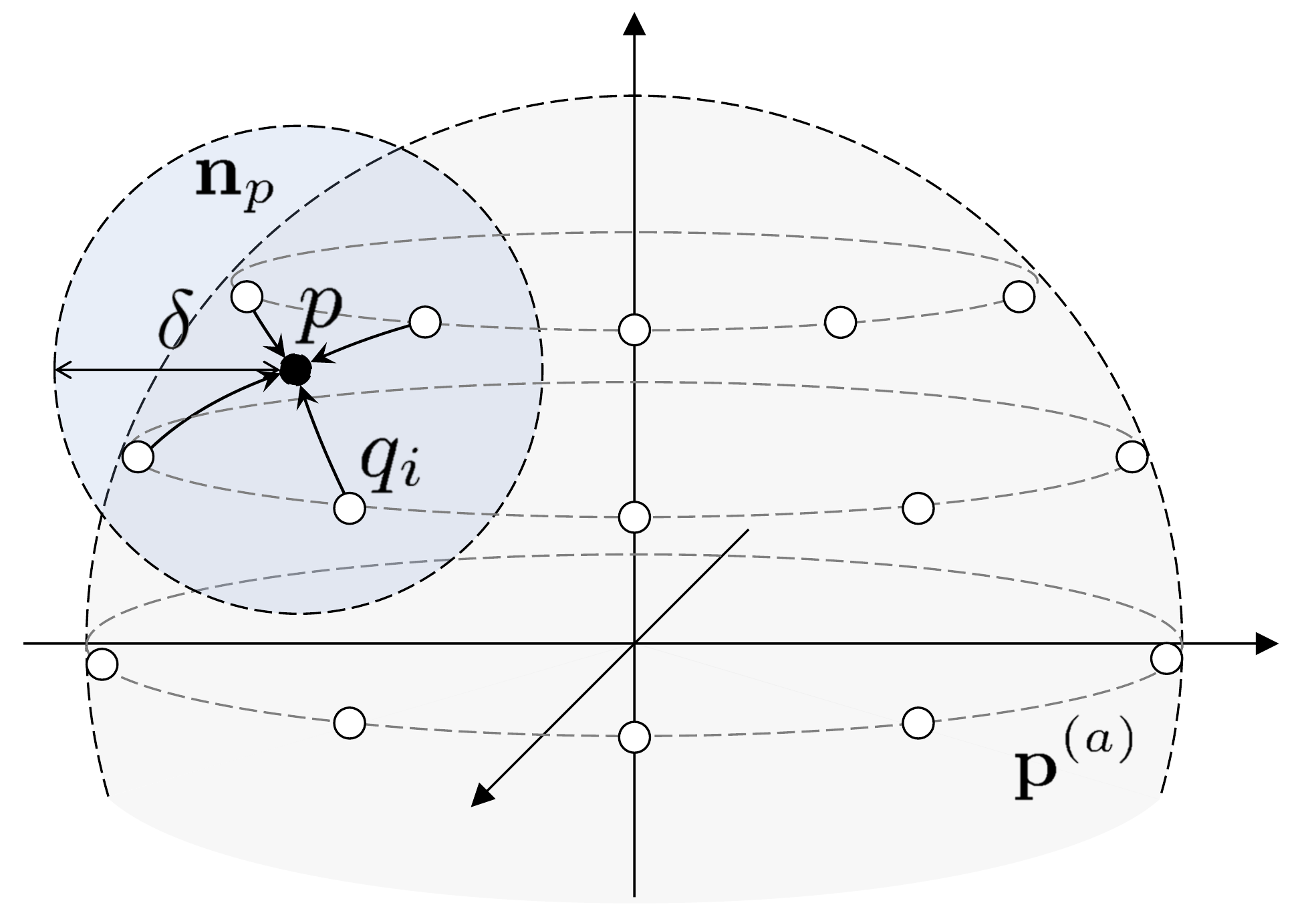}}
        \medskip
    \end{minipage}
    \caption{Schematic illustration of the proposed system.}
    \label{fig:idea}
\end{figure}

Traditionally, numerical simulations that solve the wave equation are performed to obtain HRTFs for a given head mesh.
The boundary element method (BEM) \cite{katz2001boundary,kreuzer2009fast} and finite element method (FEM) \cite{gumerov2002numerical,xiao2003finite} are commonly used.
Brinkmann et al. \cite{brinkmann2015cross} presented temporal and spectral analyses of their numerical method based on the BEM accelerated by the fast multipole method \cite{lindau2007binaural}, comparing with their measurement \cite{brinkmann2013high,otten2001factors}.
They reported that the measured and simulated HRTFs show good agreement and provided a solution to correct inaccuracies presumably caused by the mechanical setup.
Yet, precise computation of the binaural environment is computationally expensive and requires information about detailed geometries and materials of the physical environment \cite{savioja1999creating,jianjun2015natural}.

Recently, data-driven methods for binaural synthesis have been widely studied \cite{zhang2021personalized,lee2022neural,gebru2021implicit}.
These methods inherently require a sufficiently large number of data points to estimate the distribution of interest.
HUTUBS dataset \cite{brinkmann2019hutubs} is the largest among fully-public HRTF datasets \cite{guezenoc2020wide}; it provides both measured and simulated HRTFs along with anthropometric measurements.
However, the number of subjects within the HUTUBS dataset is still insufficient to split the dataset for the usual validation of the methods \cite{wang2021global}.
Therefore, in order to increase the number of training data, Chen et al. \cite{chen2019autoencoding} combined three datasets.
Furthermore, in order to combine the dataset of different coordinate systems, they applied a mapping rule, which is inapplicable for the elevation angles other than 0 \cite{chen2019autoencoding}.

To this end, we propose a method to resolve the problem of incorporating HRTF datasets of different coordinate systems.
The proposed method predicts an HRTF of the desired position from neighboring known HRTFs with their corresponding positions and anthropometric measurements.
Using subject-aware interpolation within a region of interest, our model is quantitatively and perceptually more accurate than linear interpolation.
We also show that our conditioning method advantages generalization to different coordinate systems, under limited numbers of anthropometric and positional data.

\begin{figure}
    \begin{minipage}[b]{1.0\linewidth}
        \centering
        \centerline{\includegraphics[width=0.99\linewidth]{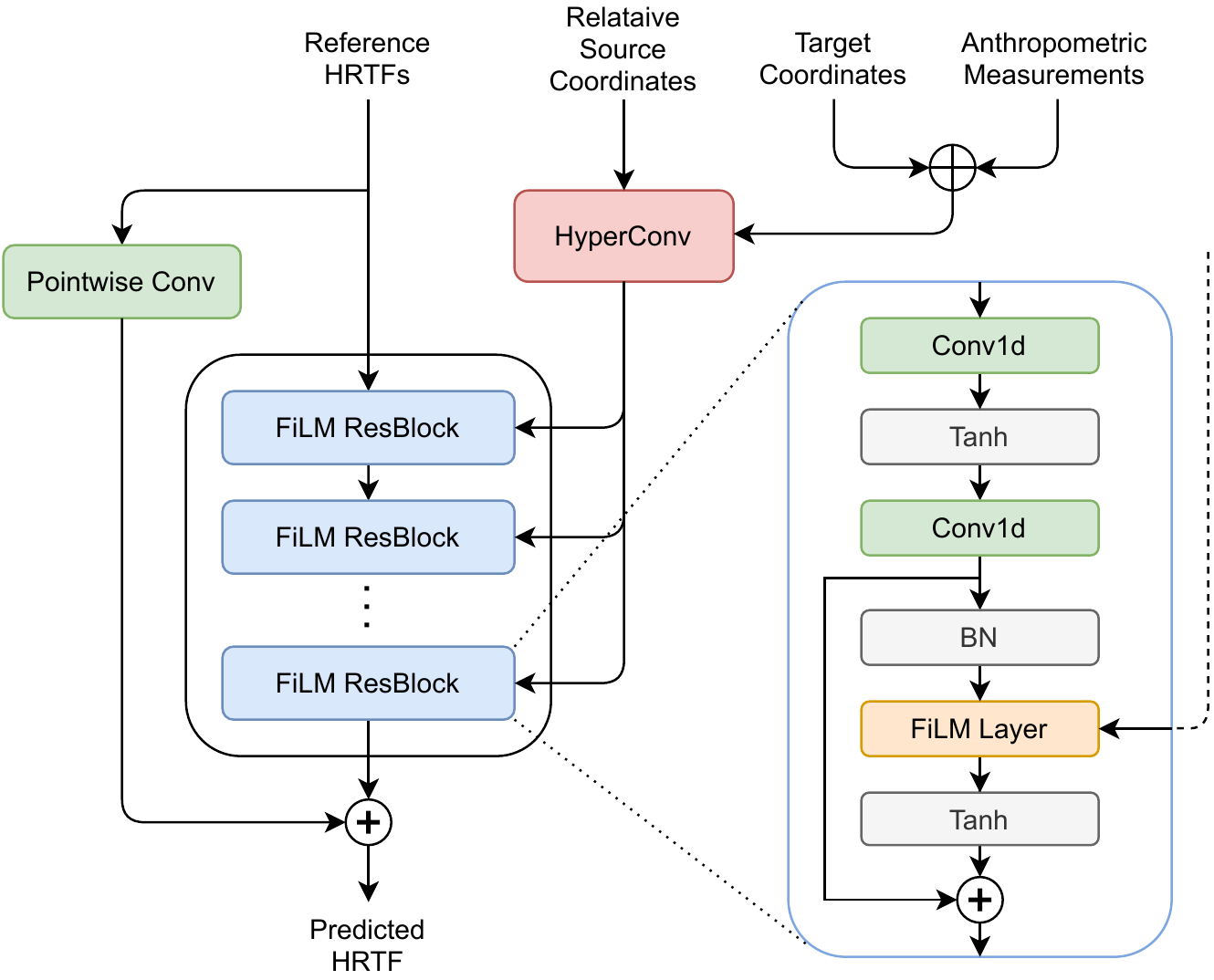}}
        \medskip
    \end{minipage}
    \caption{Block diagram of the proposed neural network architecture. The $\oplus$ operator and $+$ inside a circle denote the vector concatenation and element-wise addition, respectively.}
    \label{fig:archi}
\end{figure}

\section{Problem Statement}
\label{sec:probs}
Our main objective is to obtain HRTFs of unseen emitter positions by transforming the available reference HRTFs with a neural network conditioned on physical features. Fast Fourier Transform (FFT) of a sufficient-length HRIR represents the HRTF well, so we borrow the term ``HRTF" to denote the FFT.

Now, let $X^{(a)}_p\in\mathbb{R}^{K}$ be a HRTF with emitter position $p$ and anthropometric measurements $a\in \mathbb{R}^J$ ($K$ denotes the number of FFT frequency bins).
Furthermore, let $\mathcal{X}^{(a)}$ be a the collection of HRTF data from a subject $a$, containing responses from source points located in various positions.
Define a set of the source positions of the dataset $\mathcal{X}^{(a)}$ as
\begin{equation}
    \mathbf{p}^{(a)}=\left\{p\in\mathbb{R}^3: X^{(a)}_p\in\mathcal{X}^{(a)}\right\}
\end{equation}
represented in Cartesian coordinates.
The HRTF can either be a numerically calculated or a experimentally measured.
Assume that we aim to retrieve an HRTF from a position $p \in \mathbb{R}^3$. 
First, we define a set of neighbors of $p$ with known measurements as
\begin{equation}
    \mathbf{n}_p=\{q\in\mathbf{p}^{(a)}:0<d_E(q,p)<\delta\}
\end{equation}
where $d_E$ represents the Euclidean distance.
Then, we aim to find a function $\Phi$ which takes the neighbor measurements $X^{(a)}_{q_1},...,X^{(a)}_{q_N}$ and positions $q_1,...,q_N\in\mathbf{n}_p$ as input and estimates an HRTF of the target position $p$ as follows,
\begin{equation}
    \Phi_\theta:(X^{(a)}_q, q, p, a) \mapsto \hat{X}^{(a)}_p\in \mathbb{R}^K.
\end{equation}
Here, $q=\tilde{q}_1\oplus\cdots\oplus\tilde{q}_N$ denotes the concatenated tensors of the neighbor positions $\tilde{q}_i=q_i-p$ to the target.
$X^{(a)}_q=X^{(a)}_{q_1}\oplus\cdots\oplus X^{(a)}_{q_N}$ is the concatenated HRTFs.
The neural network parameter $\theta$ is updated to minimize the error
\begin{equation}
    \sum_{a\in\mathbf{a}}\sum_{p\in\mathbf{p}^{(a)}} \left.\mathcal{L}\left(X^{(a)}_p,\Phi_\theta(X^{(a)}_q, q, a)\right)\right\vert_{q\sim\mathbf{n}_p}
\end{equation}
with respect to the metric $\mathcal{L}$.
We may restrict the target point $p$ within the set $\mathbf{p}^{(a)}$ to solve the problem in supervised learning setting.

\section{Proposed Method}
\label{sec:method}

\subsection{Network architecture}
\label{ssec:network}
%
Our network (see Figure \ref{fig:archi}) is built upon Feature-wise Linear Modulation (FiLM) \cite{perez2018film}, an architecture designed for general-purpose conditioning.
First, we used a pointwise convolution block (\texttt{PC}) that interpolate input HRTFs by linear combination with learned coefficients.
The resultant HRTF is calibrated by the output from a stacks of FiLM residual blocks, of which we refer to as \texttt{FiLM}.
Input signals of each FiLM layer (\texttt{FiLM Layer}) are scaled and biased by learned vectors calculated from affine transformation of conditioning vectors.
In order to utilize the coordinates $p,q$, and the anthropometric measurements $a$ as a conditioning vector, we used sinusoidal encoding with the encoding dimension equal to the number of frequency bins $K$.

Yet, the number of anthropometric data within an HRTF dataset is equal to the number of subjects, it is comparably smaller than the number of coordinates, or that of the HRTFs.
Since the number of subjects are often insufficient for data-driven methods, the neural networks are likely to be overfitted to the $a$.
Therefore, other than choosing a na\"ive concatenation of $p,q$, and $a$ as the condition of \texttt{FiLM Layer}, an additional conditioning method was employed.
We introduce a new architecture that modulates the conditions for \texttt{FiLM Layer} using a hyper-convolution layer (\texttt{HyperConv}).
The \texttt{HyperConv} convolve its inputs using weight and bias tensors calculated from its conditions.
Conditioned by the anthropometric measurements and the target position, the \texttt{HyperConv} modulates the relative positions of the input HRTFs to generate conditions for the \texttt{FiLM Layer}.
We hypothesize that this conditioning framework advantages generalized prediction of HRTFs.

\noindent\textbf{Feature-wise Linear Modulation.}
The \texttt{FiLM Layer} consists of 5 stacked FiLM residual blocks \cite{perez2018film}.
We used 1D convolution with kernel size 3 instead of 2D convolution that appears in \cite{perez2018film}.
The first convolution layer of the first FiLM residual block expands the number of channels from $N$ to $4N$, and the first convolution layer of the last FiLM residual block reduce the number of channels from $4N$ to $1$.
For the other convolution layers, the number of channels remained constant.
Additionally, we used the $\tanh$ activation instead of the ReLU activation so that the training with the dB-scaled input is stable.
We formulate our feature-wise affine map for the given input signal $x_{1:K}\in\mathbb{R}^{C_{\text{in}}\times K}$ and the conditioning signal $c_{1:K}\in\mathbb{R}^{1\times K}$ as
\begin{equation}
    z_{j,:} = \mathcal{F}^{(\gamma)}(c)_{j}\cdot x_{j,:} + \mathcal{F}^{(\beta)}(c)_{j}
\end{equation}
where $\mathcal{F}^{(\gamma)}$ and $\mathcal{F}^{(\beta)}$ denote linear operators.

\noindent\textbf{Hyper-convolution.}
The hyper-convolution layer is a linear module of which the weights and biases are generated from given conditions.
In the work by \cite{richard2020neural,gebru2021implicit}, the hyper-convolution was used for conditioning the temporal convolutional network that operates in the time domain.
We adopt a similar setting with \cite{richard2020neural}, yet without dilation and different kernel size, since our \texttt{HyperConv} deals with encoded positions and anthropometric features.
Following the formulation of \cite{richard2020neural}, we express our hyper-convolution of the given input signal $x\in\mathbb{R}^{C_{\text{in}}\times K}$ and the conditioning signal $c\in\mathbb{R}^{C_{\text{cond}}\times K}$ as
\begin{equation}
    z_{:,k} = \sum_{i=0}^{\kappa-1}\left[\mathcal{H}^{(\text{w})}(c_{:,k})\right]_{:,:,i}x_{:,k+i-\lfloor\kappa/2\rfloor} + \mathcal{H}^{(\text{b})}(c_{:,k})
\end{equation}
where $\kappa$ is analogous to the kernel size of usual convolution operation, and $\mathcal{H}^{(\text{w})}$ and $\mathcal{H}^{(\text{b})}$ are small convolutional hyper-networks.

\section{Experiments}
\label{sec:experiment}

\subsection{Data}
\label{ssec:dataset}
We use HUTUBS dataset \cite{brinkmann2019hutubs} for the experiment, which consists of a total 96 number of distinct subjects.
A total of 93 number of subjects are studied by excluding three subjects (with IDs 18, 79, and 92) whose anthropometric data are unavailable.
The simulated HRTFs consist of a Lebedev grid with 1730 points. 
We transform each provided 256-sample HRIR of 44.1 kHz sampling rate into the 129-sample HRTF points via FFT.
We use 12 anthropometric features of the left ear measurements of each subject.

In order to test our model's interpolation capability for data under a coordinate system other than the Lebedev grid, we evaluate the model with the FABIAN dataset \cite{brinkmann2017high}, which uses geographical grid.
The dataset provides measured and simulated HRTFs using both head and torso models, and one of the purposes of establishing the dataset includes analysis of changes in HRTF for various head-above-torso-orientation (HATO) degrees.
Yet, we emphasize that the main feature of the FABIAN is that it provides high-resolution HRTFs, and we verify the generalization of our model for simulated HRTF with 0 HATO degree.
Each simulated HRTF of the dataset consists of 11950 source positions with 2\degree in elevation, 2\degree great circle distance in azimuth.
We also test the model performance on spatially downsampled FABIAN data, so the model reconstructs the ground truth with provided spatially downsampled HRTFs.

\subsection{Training}
\label{ssec:training}
While training our model, we uniformly sample with replacement of $N$ source coordinates for each target point $p$ from the neighborhood set $\mathbf{n}_p$.
In the test phase, we sample $N$ closest source from the target in order to assess the best achievable performance, except for the the circumstances that $|\mathbf{n}_p|$ is smaller than $N$.
To train our network, we adopt log-spectral distance (LSD) as the minimization criteria, defined by a root-mean-squared error (RMSE) between the estimated and ground-truth log magnitudes.
We employ the 5-fold cross-validation, and use AdamW \cite{reddi2019convergence} as an optimizer, with a batch size of 64.
The initial learning rate is set 0.0001 with $\beta_1=0.9$ and $\beta_2=0.999$, and the learning rate was halved if three epochs have passed without improvement.

\begin{figure}[t]
    \centering
    \centerline{\includegraphics[width=0.99\linewidth]{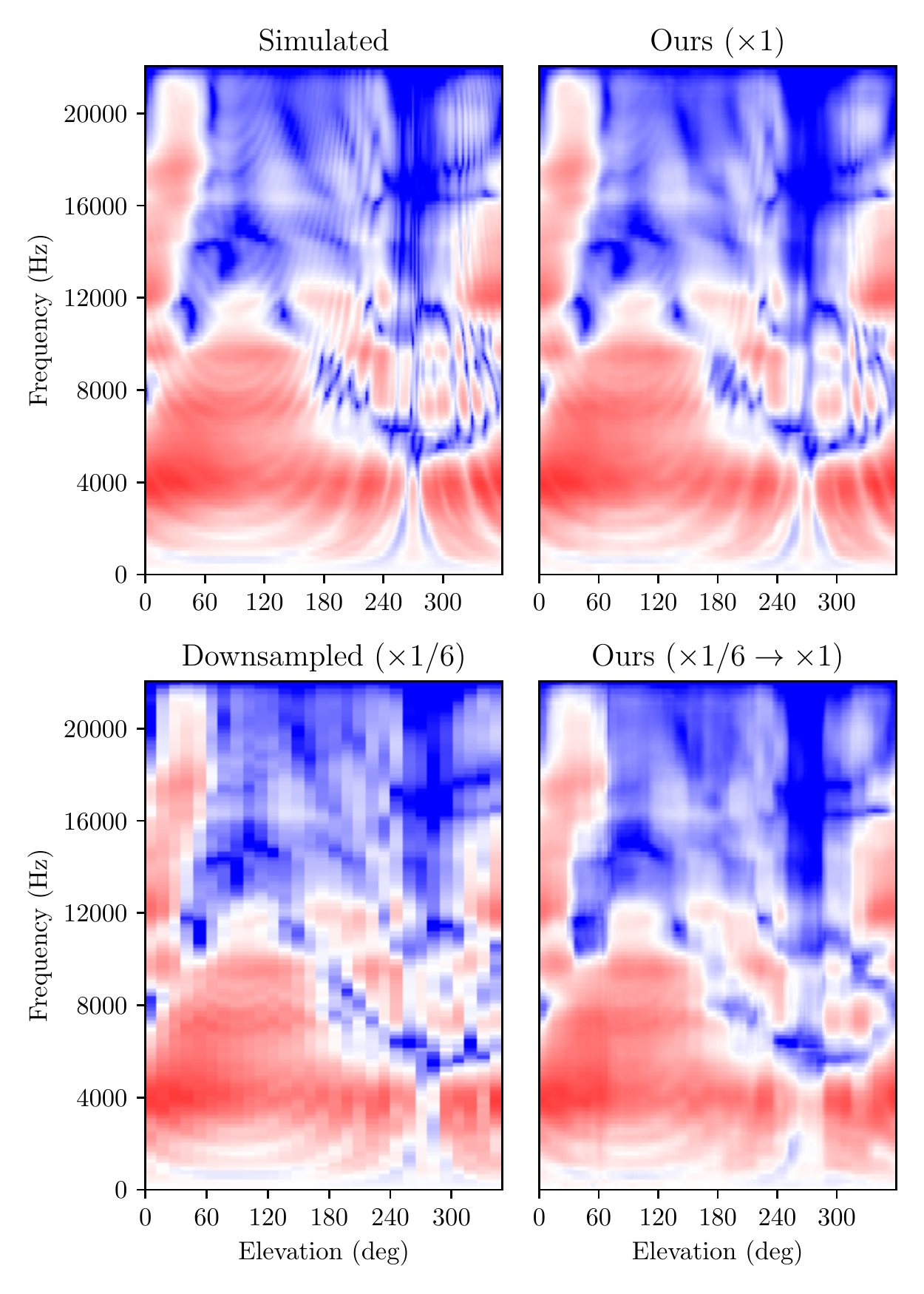}}
    \caption{Comparison between simulated HRTF of median plane of FABIAN \cite{brinkmann2017high} (\textit{top left}), estimated output of our model (\textit{top right}), simulated HRTF of median plane of FABIAN downsampled by a factor of $\frac{1}{6}$ (\textit{bottom left}), and reconstruction by our model with provided the downsampled data (\textit{bottom right}).}
    \label{fig:fabian-interp}
\end{figure}

\begin{table}[b]
    \centering
    \begin{tabular}{clc}
        \toprule
        & Model & LSD \\
        \midrule
        (a) & \texttt{PC} & 1.875 \\
        (b) & \texttt{PC} + \texttt{FiLM} & 1.767 \\
        (c1) & \texttt{PC} + \texttt{FiLM} + \texttt{FiLM Layer} & 1.767 \\
        (c2) & \texttt{PC} + \texttt{FiLM} +  \texttt{HyperConv} & \textbf{1.736} \\
        \bottomrule
    \end{tabular}
    \caption{Comparison of average log-spectral distortion for all directions depending on the modules used for training.}
    \label{tab:ablation}
\end{table}

\subsection{Ablation study}
\label{ssec:ablation}
We conduct an ablation study on our best model to understand how the network learns interpolation by conditioning the physical features.
We report the models with $N=8$ in $\delta=0.3$ on HUTUBS in Table \ref{tab:ablation}.
Beginning from the case of using only the pointwise convolution (\texttt{PC}) as a baseline, we add the FiLM residual network (\texttt{FiLM}), or additionally use FiLM layer (\texttt{FiLM Layer}) or hyper-convolution layer (\texttt{HyperConv}) in order to condition \texttt{FiLM}.
The baseline (a) is equivalent to a linear operation that superposition the input HRTFs with learned coefficients.
No conditioning method is adopted in the baseline.
For (b), we use similar architecture of FiLM network as proposed in \cite{perez2018film}, and conditioned the concatenated vectors of encoded positions and anthropometric features.
We further hypothesize that the condition which inform the positions of input HRTFs should be modulated by the position of target HRTF and anthropometric features, as illustrated by red box in Figure \ref{fig:archi}.
Table \ref{tab:ablation} (c1) and (c2) show the result of using different conditioning methods for modulating the conditions.
In (c1), FiLM residual block (single blue box in Figure \ref{fig:archi}) with kernel size 1 is utilized for modulating the conditions.
Using the conditioning method as (c2) advantages the most accurate interpolation of HRTFs.
We empirically observe that hyper-convolution advantages conditioning the physical features in predicting HRTFs.

\begin{figure}[b]
    \begin{minipage}[b]{0.49\linewidth}
        \centerline{\includegraphics[width=\linewidth]{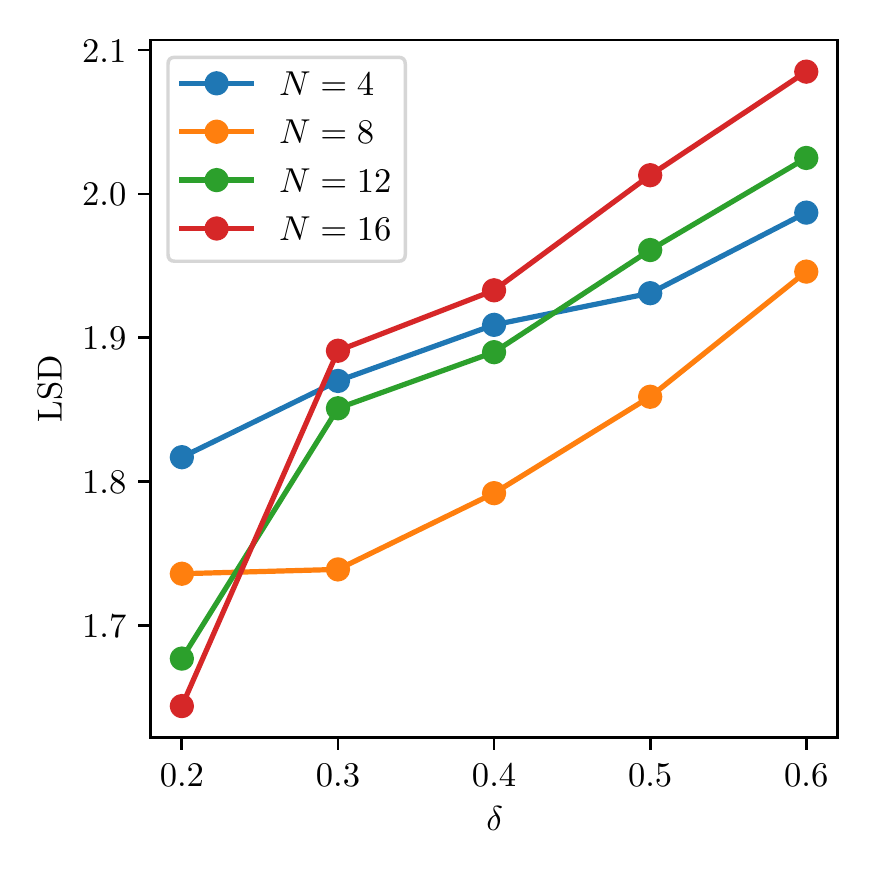}}
        \centerline{(a) Varying $\delta$ with fixed $N$.}\medskip
    \end{minipage}
    \hfill
    \begin{minipage}[b]{0.49\linewidth}
        \centerline{\includegraphics[width=\linewidth]{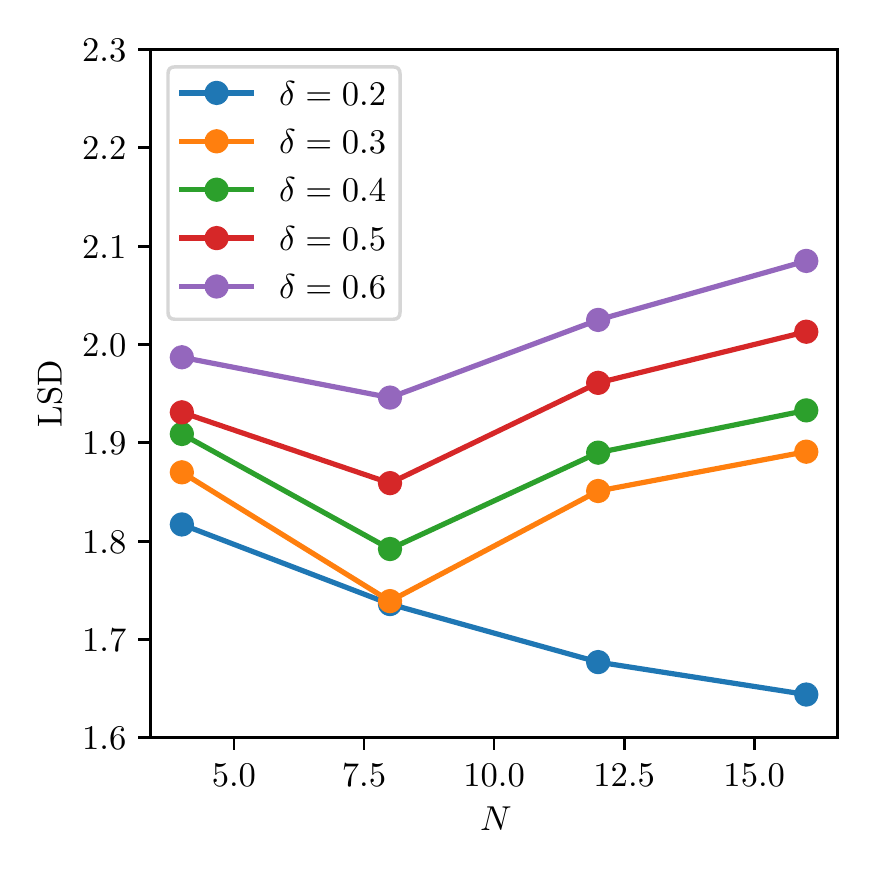}}
        \centerline{(b) Varying $N$ with fixed $\delta$.}\medskip
    \end{minipage}
    \caption{Comparison of model performance for the radius $\delta$ and the number of sampling points $N$ within the neighborhood $\mathbf{n}_p$.}
    \label{fig:n-study}
\end{figure}

\subsection{Results}
\label{ssec:result}
We compare our model with a baseline which linear-interpolates the two closest samples within a plane of equal azimuth or elevation.
We denote the horizontal/median/frontal plane by `Hor.'/`Med.'/`Fro.', respectively.
Average result for the samples of every azimuth/elevation angle is denoted by `All'.

\begin{table}[t]
    \centering
    \begin{tabular}{ccccc}
        \toprule
        Direction & All & Hor. & Med. & Fro. \\
        \midrule
        Linear Interp.      & 6.538 & 6.418 & 2.289 & 6.842 \\
        Ours, $\delta=0.2$  & \textbf{1.761} & \textbf{2.093} & \textbf{1.139} & \textbf{2.198} \\
        \bottomrule
    \end{tabular}
    \caption{Comparison of average LSD on test set of HUTUBS.}
    \label{tab:hutubs-test}
\end{table}
\noindent\textbf{Interpolation.}
Table \ref{tab:hutubs-test} compares average performance of our model with $N=16$ and $\delta=0.2$ with the linear interpolation baseline.
Our method outperforms the baseline for all directions and every planes.
In the Lebedev grid, it is difficult to find planes that make the distances between points densely distributed.
Because of this, linear interpolation is easy to fail in HUTUBS.
On the other hand, linear interpolation of the FABIAN dataset \cite{brinkmann2017high} shows comparably better performance than the linear interpolation result in HUTUBS.
Table \ref{tab:recon-fabian} shows the performance of our model, which was trained using HUTUBS and tested using simulated HRTF in FABIAN.
Although the tested HRTFs are aligned in unseen coordinate system and radius, our model performed similar to the result observed in HUTUBS.
The model also outperforms the baseline for horizontal and median planes, and the LSD averaged for all directions.

\begin{table}[t]
    \centering
    \begin{tabular}{ccccc}
        \toprule
        Direction & All & Hor. & Med. & Fro. \\
        \midrule
        Linear Interp. & 2.539 & 3.145 & 3.245 & \textbf{2.071} \\
        Ours, $\delta=0.6$ & \textbf{1.897} & \textbf{2.190} & \textbf{1.686} & 2.514 \\
        \midrule
        Ours ($\times1/2$)    & 2.105 & 2.481 & 2.414  & 3.000  \\
        Ours ($\times1/4$)    & 2.353 & 2.806 & 2.775 & 3.280 \\
        Ours ($\times1/6$)    & 2.513 & 2.941 & 2.861 & 3.269 \\
        \bottomrule
    \end{tabular}
    \caption{Comparison of average LSD tested using FABIAN dataset.}
    \label{tab:recon-fabian}
\end{table}
\noindent\textbf{Super-resolution.}
We also test our model's interpolation performance under more spatially downsampled HRTF grids by factor of $1/T$ (higher $T$ implies harder super-resolution task).
In order to assure the existence of input points within the neighborhood $\mathbf{n}_p$, we show the performance of our model trained using $N=16$ within $\delta=0.6$, even though the model is not the best one of our analysis.
Table \ref{tab:recon-fabian} shows average LSD of our model with provided reference HRTFs uniformly coarsened by factor of $1/2$, $1/4$, or $1/6$.
Although given only $1/6$ times less data compared to the original, our model maintains generally better performance than the baseline.
For the plotted results of super-resolution task, we refer to Figure \ref{fig:fabian-interp}.
We exclude the measured HRTFs of FABIAN dataset from the experiment since we cannot sample enough $N$ number of measured FABIAN HRTFs within the $\delta$ of our model.

\noindent\textbf{Perceptual evaluation.}
Table \ref{tab:pesq-fabian} compares perceptual evaluation of speech quality (PESQ) \cite{rec2005p} of speech samples rendered with the HRTFs.
PESQ is a perceptual metric widely used especially in speech enhancement, or binaural speech synthesis \cite{leng2022binauralgrad}.
We measure PESQ between the speech rendered with the simulated FABIAN HRTFs and those rendered using the estimated HRTFs.
For the speech samples we use a subset of VCTK \cite{yamagishi2019cstr} corpus which includes speech recordings with 48 kHz sampling rate uttered by 110 English speakers with various accents.
We downsample the speech data into rate of 44.1 kHz to render the HRTFs.
The rendered speeches are downsampled into rate of 16 kHz to measure the PESQ.
Table \ref{tab:pesq-fabian} implies that the speech samples rendered using our model's HRTFs are perceptually similar to the speech rendered using the simulated HRTFs, outperforming the baseline.
In contrast to linear interpolation, where PESQ is reduced in more difficult super-resolution scenarios (higher $T$), our model shows robust PESQ scores.

\noindent\textbf{Setting neighborhood.}
Figure \ref{fig:n-study} shows the average performance of the models trained with different neighborhood constraints.
We explore appropriate size $\delta$ and number of sampling points $N$ of the neighborhood $\mathbf{n}_p$ using the test set of HUTUBS.
The LSD increases for all models with increment in the neighborhood radius $\delta$ when the number of sampling points is fixed.
For sufficiently small $\delta$, the error decreases when more points were sampled.
However, the models trained with $N=8$ consistently performs better than the others for $\delta\geq0.3$.
Models trained with larger $\delta$ get less chance to be provided with necessary cues in predicting the HRTFs of target position while training.
Thus training models with larger $\delta$ involve higher difficulties of optimizing the model.
However, we empirically observe that the models trained using larger $\delta$ generalized better, especially for upsampling with large scales.

\begin{table}[t]
    \centering
    \begin{tabular}{ccccc}
        \toprule
        Downsample factor & $T=1$ & $T=2$ & $T=4$ & $T=6$ \\
        \midrule
        Linear Interp. & 3.47 & 3.47 & 3.30 & 3.33 \\
        Ours, $\delta=0.6$ & \textbf{4.10} & \textbf{4.11} & \textbf{4.13} & \textbf{4.13} \\
        \bottomrule
    \end{tabular}
    \caption{Comparison of PESQ (the higher the better) for HRTF-rendered speech samples.
    We estimate FABIAN HRTFs for every 11950 source positions of the database upon the spatially downsampled HRTF grids.
    The HRTFs are rendered to the speech to measure the perceptual distance with the simulated HRTFs.
    }
    \label{tab:pesq-fabian}
\end{table}

\section{Conclusion}
\label{sec:conclusion}
In this work, we introduce a neural network that interpolates given HRTFs by conditioning source positions and anthropometric measurements. 
We explore the use of two conditioning methods, FiLM and hyper-convolution, for our network that estimate the HRTF of the desired position for given HRTFs sampled from its neighborhood.
Quantitative analysis shows that our method precisely interpolates the HRTFs, outperforming the baseline.
The ablation experiment shows that the proposed conditioning method performed better than the rest of the experimented methods.
Also, the model with the proposed conditioning scheme generalizes well to the unseen dataset with the unseen grid system.
Furthermore, we demonstrate that our network is able to reconstruct the spatially downsampled HRTFs in both quantitative and perceptual measures.


\newpage
\bibliographystyle{IEEEbib}
\bibliography{refs}

\end{document}